# Speckle-driven single-shot orbital angular momentum recognition with ultra-low sampling density

Zhiyuan Wang[1,2,3,8], Haoran Li[1,2,3,8], Tianting Zhong[4,5], Qi Zhao[3,4], Vinu R. V[1,2], Huanhao Li[3,4], Zhipeng Yu[3,4], Jixiong Pu[1,2], Ziyang Chen[1,2] ✉, Xiaocong Yuan[6] ✉ & Puxiang Lai[3,4,7] ✉

Orbital angular momentum (OAM) recognition of vortex beams is critical for applications ranging from optical communications to quantum technologies. However, conventional approaches designed for free-space propagation struggle when light passes through scattering media, such as multimode fibers (MMF), and often rely on high-resolution sensors with tens of thousands of pixels to record detailed intensity profiles. Here, by harnessing scattering media as intrinsic encoders rather than detrimental factors, we introduce a speckle-driven OAM recognition technique termed spatially multiplexed points detection (SMPD). This method extracts intensity information from a few spatially distributed points in a speckle plane, where object feature is naturally multiplexed, thereby transforming scattering from a detrimental effect into an efficient encoding mechanism while drastically reducing sampling requirements. Remarkably, it achieves over 99% retrieval accuracy for OAMs recognition using just 16 sampling points, corresponding to a sampling density of 0.024% compared with conventional imaging-based approaches. Furthermore, spatiotemporally interleaved vortex beams decoding, high-capacity OAM-multiplexed communication, MNIST, and Fashion-MNIST classification are implemented to verify the versatility of SMPD. This work establishes a scalable strategy for efficient optical information processing and fiber-based sensing in complex environments.

The precise manipulation of light's orbital angular momentum (OAM) has emerged as a cornerstone of modern optics[1–4], unlocking an infinite-dimensional basis for applications ranging from high-capacity optical communication[5–12] and quantum information protocols[13–15] to multidimensional sensing[16–18]. Central to this paradigm is the vortex beams, whose helical phase structure, characterized by an azimuthal phase dependence of $\exp(i\ell\theta)$, enables robust encoding of information into discrete OAM modes. These modes form the backbone of spatial multiplexing strategies that transcend classical transmission limits[1–3].

To decode OAM modes, approaches such as interferometry[19–21], vortex diffraction grating[22–24], and machine learning[25,26] have proven effective in free-space environments. While demonstrating success, they generally depend on high-resolution imaging to capture well-defined full-field light structures, necessitating thousands of pixels or more and computationally intensive processing. Such reliance on dense spatial sampling imposes stringent trade-offs; it throttles system bandwidth, complicates scalability, and obstructs real-time operation[27,28]. Consequently, applications demanding compact form factors, high-speed detection, or operation in photon-limited scenarios, cornerstones of next-generation photonic technologies, remain largely out of reach.

[1]Huaqiao University, College of Information Science and Engineering, Xiamen, China. [2]Huaqiao University, Fujian Provincial Key Laboratory of Light Propagation and Transformation, Xiamen, China. [3]Hong Kong Polytechnic University, Department of Biomedical Engineering, Hong Kong, China. [4]Hong Kong Polytechnic University, Shenzhen Research Institute, Shenzhen, China. [5]The University of Hong Kong, Department of Electrical and Electronic Engineering, Hong Kong, China. [6]Shenzhen University, Nanophotonics Research Center, Shenzhen Key Laboratory of Micro-scale Optical Information Technology, Institute of Microscale Optoelectronics, Shenzhen, China. [7]Hong Kong Polytechnic University, Photonics Research Institute, Hong Kong, China. [8]These authors contributed equally: Zhiyuan Wang, Haoran Li. ✉e-mail: ziyang@hqu.edu.cn; xcyuan@szu.edu.cn; puxiang.lai@polyu.edu.hk

Recent advances reveal that the ostensibly disordered nature of speckle fields marks a profound redundancy in their information content[29–31]. Unlike direct imaging systems, which localize spatial information to discrete regions, multiple scattering in complex media redistributes input-field contributions across the entire speckle pattern (see Supplementary Note 1 for more details). This non-local encoding enables information retrieval based on partial-field sampling for tasks such as imaging through scattering media[32], diffuser-based super-resolution imaging via sub-Nyquist sampling[33], and OAM recognition[5,9,34–36] from arbitrary speckle subregions.

Yet, leveraging this redundancy for ultralow-sampling detection remains underexplored. Although indeed utilizing a portion of the speckle field, current implementations still often rely on capturing a high-resolution intensity profile pattern (comprising thousands to millions of data points). A pivot challenge persists: What constitutes the minimal spatial sampling required to maintain accurate, scalable mode discrimination?

To address this, we propose spatially multiplexed points detection (SMPD), a framework for single-shot pattern classification with unprecedented sampling efficiency. Instead of recording full-field speckles, SMPD spatially samples intensity fluctuations at a sparse array of spatially multiplexed single-pixel detectors (SPDs), each capturing globally encoded information from the scattered field. A custom-designed neural network deciphers these sub-Nyquist intensity sequences, achieving robust OAM recognition with a sampling density of 0.024%, 4096 times lower than conventional imaging-based methods, while maintaining over 99% accuracy in diversity OAM recognition scenarios. Beyond OAM-mode retrieval, SMPD generalizes to classify mode-dependent features and distinguish complex optical fields, including handwritten digit recognition, grayscale images of clothing classification, and axicon angle identification, demonstrating its versatility across diverse scenarios.

By refining the spatial sampling-information capability trade-off, SMPD transcends the limitation of imaging-centric detection paradigms. This approach enables high-dimensional optical sensing and communication systems that are both compact and bandwidth-efficient, critical for real-time operation in photon-starved environments. Furthermore, its ultralow sampling requirements may unlock applications for non-line-of-sight (NLOS) sensing, low-power LIDAR, and quantum-enhanced sensing, where conventional pixel-dense sensors, constrained by their limited bandwidth and spectral ranges, prove impractical. SMPD thus establishes a pathway toward scalable photonic technologies, bridging the gap between theoretical information redundancy and practical, resource-efficient detection.

## Results

The proposed SMPD framework was rigorously evaluated through a series of experiments to validate its efficiency in various complex scenarios of OAM recognition and its applicability to pattern classification. Below, we present the key findings that highlight the method's performance, robustness, and scalability.

### High-accuracy OAMs recognition with ultra-low sampling density

Figure 1 illustrates the concept of SMPD, where two orthogonally polarized vortex beams transmitted through a multimode fiber (MMF) generate a speckle pattern at the distal end of the fiber. Instead of conventional full-field imaging, SMPD employs spatially distributed SPDs to sample sparse intensity points. These measurements are processed by a customized neural network termed as Recognizing OAM Artificial Neural Network (ROAM-ANN), which secures two paths to extract the features of the inputs in different dimensions and eventually outputs the encoded OAMs from the pointwise intensity sequences.

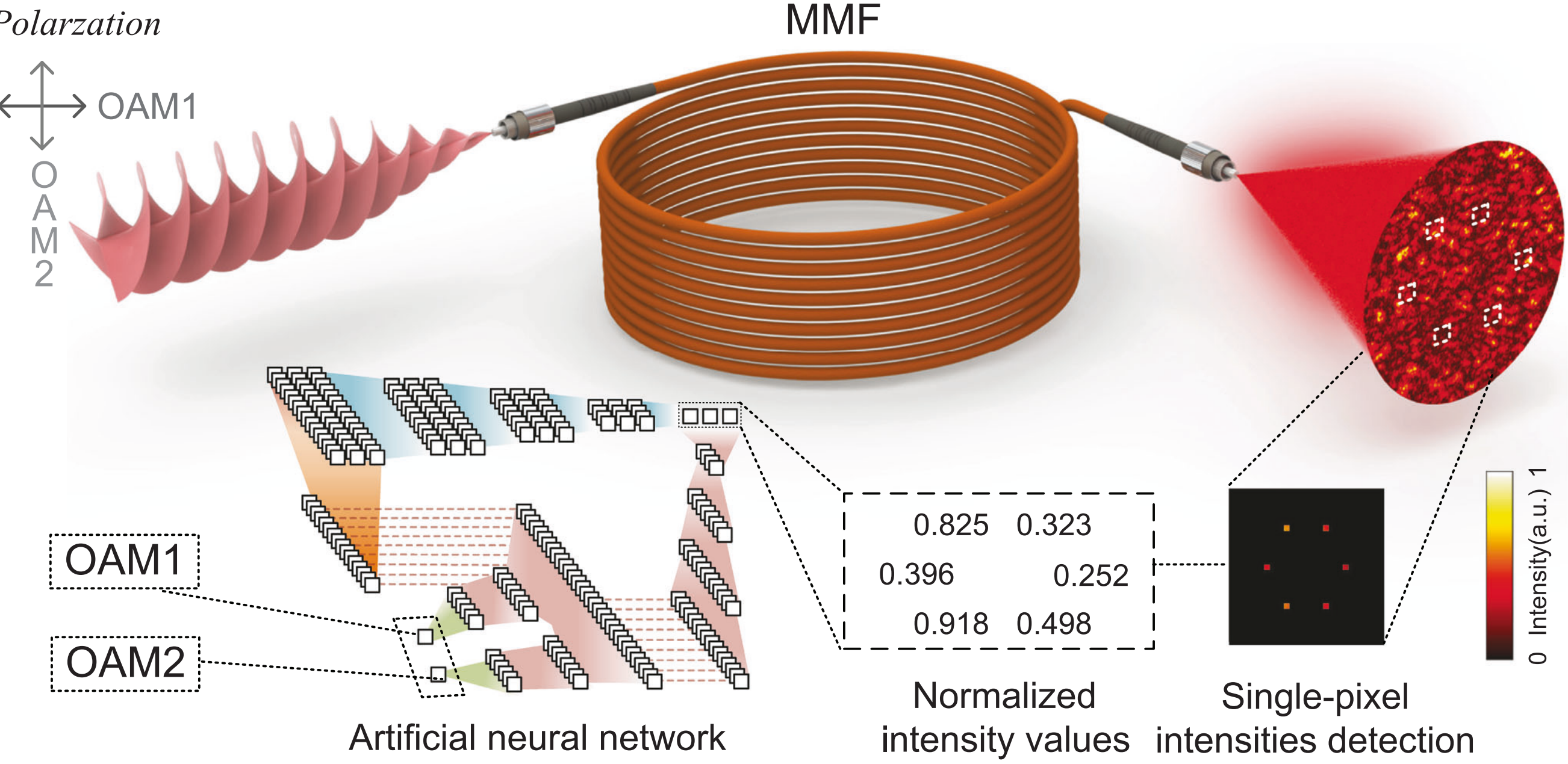

**Fig. 1 | SMPD framework for single-shot OAM recognition via ultra-sparse speckle sampling.** Two orthogonally polarized vortex beams propagate through a multimode fiber (MMF), generating a scrambled speckle pattern at the distal end. The intensity values are sampled by several strategically distributed single-pixel detectors in the speckle field and input into a designed neural network responsible for recognizing the OAMs of the input vortex beams.

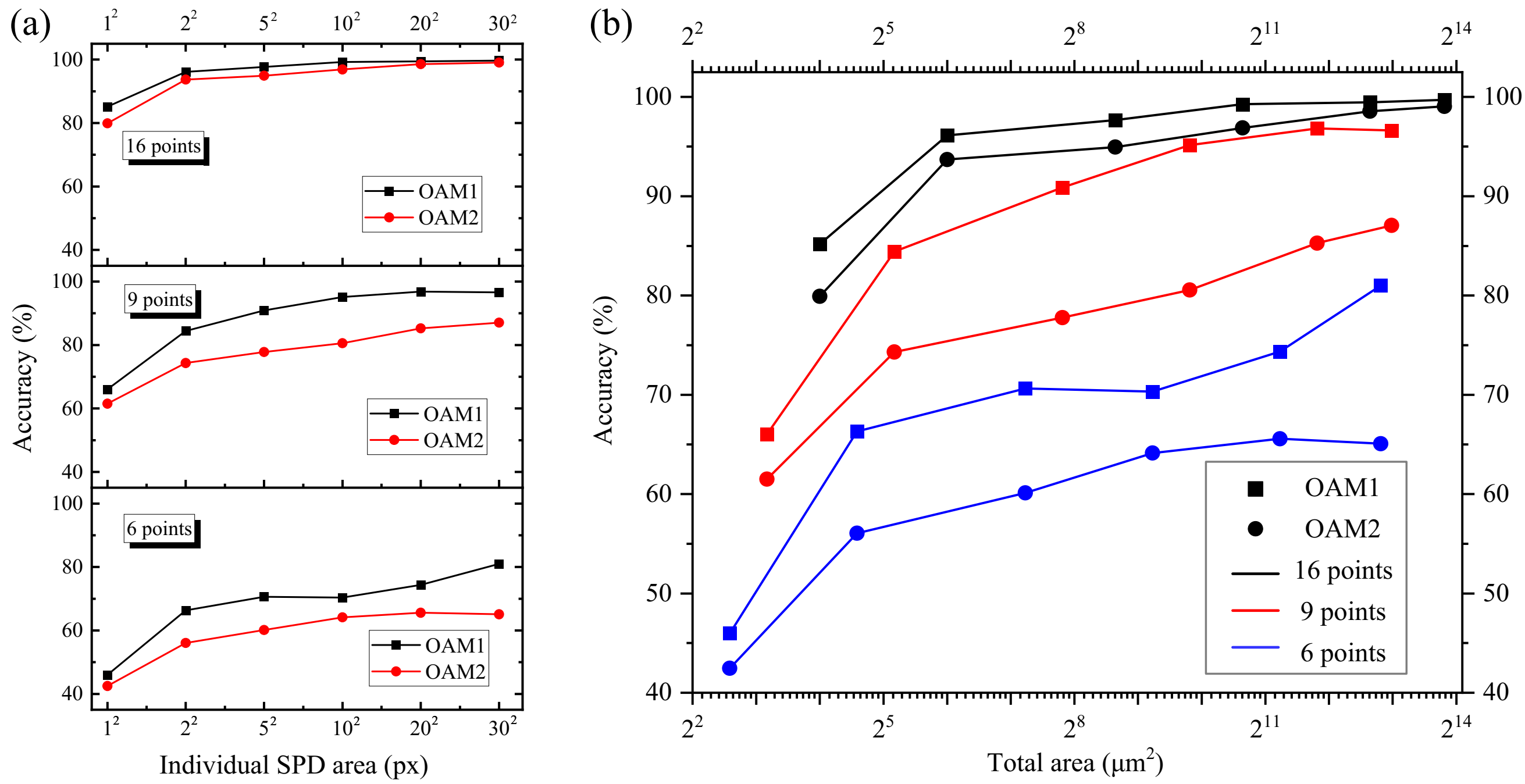

**Fig. 2 | Recognition accuracies of single pixel detector (SPD) configurations on the test set. a** Impact of SPD size and count: Recognition accuracy for OAM1 and OAM2 modes ($\ell = 1$–10.9) versus individual SPD area (1 × 1 px to 30 × 30 px) and number of SPDs (6–16). **b** Sampling efficiency analysis: Accuracy versus total effective detection area.

To quantify the impact of SPD parameters on recognition accuracy, we systematically varied the number and size of SPDs (Fig. 2). A charge-coupled device (CCD) camera with various masks was used to create different SPDs of varying numbers and sizes. Here, SPD is defined in units of 'px', and 1 px corresponds to the area of a single element ($7.4^2$ μm$^2$) in the CCD camera. For instance, an SPD area of $5^2$ px indicates 5 by 5 neighboring CCD elements are combined to form one SPD, and a total of 9 points means that 9 such SPDs are imposed on the detecting plane via the mask. For each SPD, the light intensities detected by CCD's light-sensitive elements are accumulated, and the recorded intensity sequence of the SPD array is used as the input of the network.

The recognition results for the test sets, each containing 1000 intensity sequences randomly selected from a dataset of 10,000 combinations of two orthogonally polarized vortex beams (both OAMs varying from 1 to 10.9 with an interval of 0.1) and excluded from the training process, are shown in Fig. 2a (more detailed results can be found in Supplementary Note 2). As seen, 16 SPDs (each spanning 30 × 30 px) achieved >99% accuracy for both OAM modes. Critically, the total sampling density of this configuration is approximately 0.024% (16 vs. 256 × 256), corresponding to 4096 times lower than that of conventional imaging-based methods[9] that usually require 256 × 256 or more pixel arrays. Even with SPDs reduced to 2 × 2 px (retaining 16 detectors), accuracy remained above 93%.

That said, a clear trade-off emerged between SPD count and SPD size: reducing either the number of sampling points (which is critical for capturing multiplexed information and speckle correlations) or the sensing area (which helps resist perturbations and avoids sampling low-intensity or uninformative zones, as further discussed in Supplementary Note 1.4) led to degraded performance. For instance, with only 6 SPDs (1 × 1 px), accuracy dropped below 50%, highlighting the necessity of using sufficient sampling points and an adequately large sensing area to ensure robust performance, mitigate noise, and reduce sensitivity to pixel displacement. To better visualize this trend, Fig. 2b plots recognition accuracy as a function of the total SPD area across different measurement sets.

In addition to recognizing orthogonally polarized vortex beams, we also conducted simulations involving pairs of linearly polarized vortex beams with varying polarization angle differences (see details in Supplementary Note 6). The proposed method successfully distinguished overlapping vortex beams with polarization angle differences as small as 5°, achieving a recognition accuracy of approximately 90%. As the angular separation decreased to 3°, the recognition accuracy declined to around 80%, reflecting the increasing challenge of resolving nearly co-polarized beams. These results further demonstrate the sensitivity and robustness of our method in scenarios involving subtle polarization differences. It indicates that multiplexing OAM inputs across different polarization states in polarization space, combined with the proposed SMPD method, can largely reduce the amount of stored data while significantly improving the channel capacity of OAM transmission. This approach enables the potential for ultra-large capacity OAM transmission.

### Spatial deployment flexibility

A key advantage of SMPD lies in its insensitivity to the spatial deployment of SPDs. As shown in Table 1, three distinct spatial deployments of 9 SPDs (with a fixed total area of 10× 10 px) yielded minimal accuracy fluctuations: 89.40–95.15% for OAM1 and 78.05-80.55% for OAM2. This flexibility arises from the non-local encoding of information inherent to speckle patterns, where each SPD integrates contributions from the entire input field.

While our experimental results indicate that more sparsely distributed sampling regions yield slightly higher reconstruction accuracy than densely clustered ones, despite having the same number of sampling points and total area, this advantage is not absolute. Specifically, under conditions of strong scattering, where the information content can be fully scrambled and uniformly distributed across the speckle field, the choice of sampling region configuration (dense or sparse) has minimal effect on recognition accuracy (see details in Supplementary Note 1.4).

However, in practical scenarios, such as with speckles generated by a small-core MMF, a strong scrambled condition is difficult to achieve. In these cases, the speckle patterns exhibit larger grain sizes, leading to high redundancy in proximal sampling. Sparse sampling, by contrast, collects information from more uncorrelated regions of the speckle field and thus significantly improves reconstruction accuracy. Crucially, sparse sampling shows robust performance across different scattering regimes. It remains effective whether the speckle grains are large (indicative of insufficient scattering) or small (indicative of strong scattering), offering broad applicability and better practical performance under varying physical conditions.

**Table 1 | Recognition with different spatial deployments of SPDs (fixed total area and count)**

| Spatial allocations of 9 SPDs | | | |
|---|---|---|---|
| Recognition accuracy of OAM1 (%) | 89.40 | 95.15 | 93.85 |
| Recognition accuracy of OAM2 (%) | 78.05 | 80.55 | 79.70 |

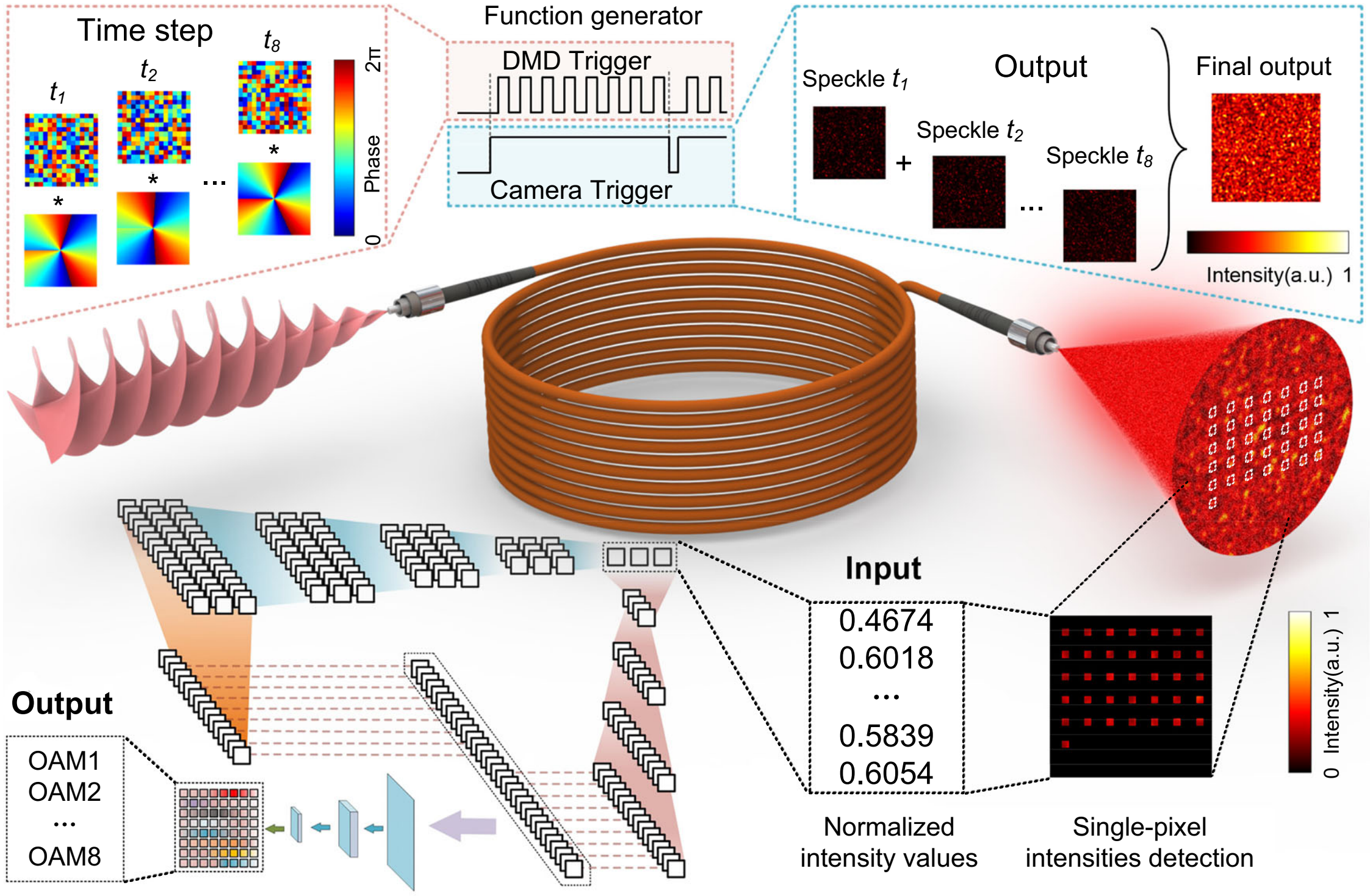

**Fig. 3 | Schematic of the encoding and decoding process for spatiotemporally interleaved vortex beams transmission.** Eight vortex beams with different OAM values are sequentially transmitted through the MMF. The charge couple device (CCD) camera records a single speckle image, representing the spatiotemporally superposition of all beams, synchronized with a digital mirror device (DMD) and a function generator. A sampling mask is applied, and the sampled intensities are input into a modified ROAM-ANN to decode the OAM information.

## Spatiotemporally interleaved OAMs decoding via SMPD

Figure 3 illustrates the conceptual framework of the experiment (see Supplementary Note 3 for details). Each vortex beam, carrying a randomly selected OAM value alongside its corresponding temporal phase masks, is transmitted through the MMF in chronological order. At the distal end of the MMF, a CCD camera captures a single speckle image representing the superposition of the eight individual speckle patterns. This process is synchronized using a function generator, which triggers the digital micromirror device (DMD) eight times and the camera once during each acquisition cycle.

To enable single-pixel detection, eight masks with varying point densities (ranging from 20 to 49 points) are applied to the speckle patterns, effectively serving as virtual SPDs arrays, with each SPD covering an area of 10 × 10 pixels. Notably, the densest mask (7 × 7 SPD array, 49 sampling points) corresponds to just 0.07% of the 256 × 256 speckle image. Using this configuration, 20,000 speckle-label pairs were collected, each containing eight OAM values randomly selected from the 2.1–2.8 range. These data were split into training, validation, and testing sets in an 8:1:1 ratio, and processed through a modified ROAM-ANN (see Supplementary Note 4).

Figure 4 presents recognition performance under various sampling conditions, reporting both collective and individual OAM detection accuracy, as well as test loss. Collective recognition accuracy refers to the success rate of correctly identifying all OAM values within a single multi-OAM input. In this context, an input consists of eight distinct OAM modes, and a prediction is considered correct only if all eight are accurately classified. Using 49 sampling points (i.e., a 7 × 7 SPD array), the collective recognition accuracy reaches 98.6%, indicating highly reliable multi-mode detection. Individual recognition accuracy, on the other hand, measures the accuracy of identifying each OAM value independently, regardless of the correctness of other values in the same input. Under the same 49-point sampling configuration, the individual recognition accuracy achieves 99.82%, with only 28 misclassifications out of 16,000 individual OAM values in the test set. Even under more constrained conditions, using only 20 sampling points, the method maintains a robust individual recognition accuracy of approximately 85%, demonstrating strong performance even with limited spatial information.

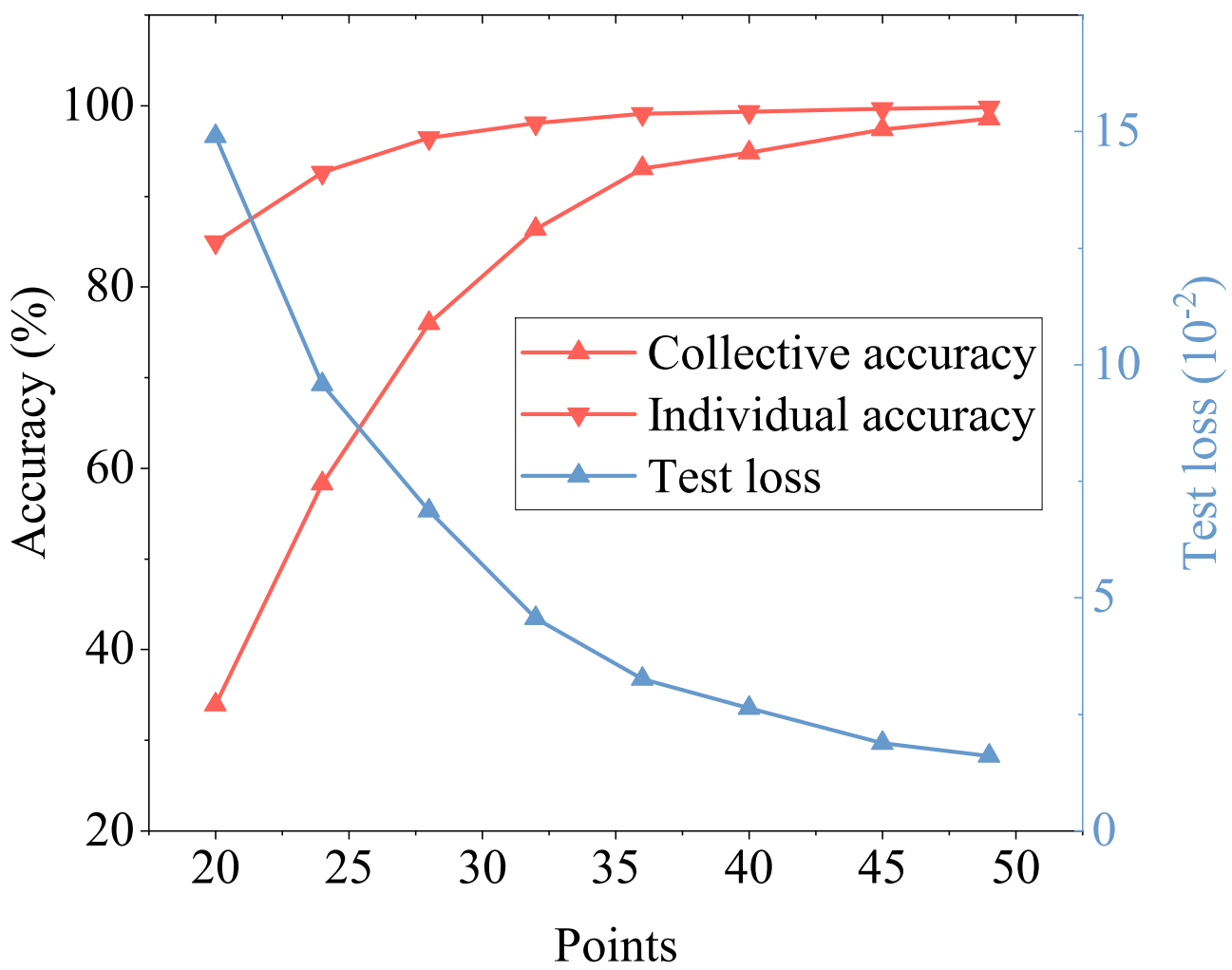

**Fig. 4 | Recognition accuracy and test loss vs. number of sampling points.** Collective accuracy: the proportion of test samples in which all OAM modes are correctly identified within a single multi-mode input; Individual accuracy: correctness of each OAM mode, evaluated independently.

### High-capacity optical communication via SMPD

To demonstrate real-world applicability, we implemented an OAM-multiplexed optical communication system (Fig. 5). Two spiral phase images (100 × 100 pixels, 8-bit color depth) were encoded using OAM bases ranging from 2.1 ~ 2.8 and multiplexed via the optimized Lee hologram[37] (see Supplementary Note 5 for details). A DMD was employed to effectively generate multiplexed vortex beams carrying the encoded data, which were then coupled into the MMF. As the beams propagated through the MMF, speckle patterns formed at the distal end. At the receiver, SMPD decoded speckle patterns using three configurations: 16, 9, or 4 SPDs. With 16 SPDs (10 × 10 px each), decoding error rates remained <0.2% (Fig. 5c). Reducing SPD size to 1 × 1 px slightly increased errors (due to localized noise and pixel displacement) but maintained robust performance. Even with only 4 SPDs (10 × 10 px each), the system achieved 74% accuracy, underscoring SMPD's resilience under extreme under-sampling. This experiment validates SMPD's potential for high-speed, low-bandwidth optical communication, particularly in resource-limited settings.

### Generalizable pattern recognition via ultra-sparse speckle sampling

To demonstrate SMPD's versatility beyond OAM retrieval, we extended its application to handwritten-digit and clothing classification using the MNIST dataset(Binary images)[38] and Fashion-MNIST dataset(8-bit grayscale images)[39].

Conventional speckle-based digit classification relies on high-resolution speckle imaging and computationally intensive spatial feature extraction. In contrast, as illustrated in Fig. 6a, SMPD employed simulated scattering transmission of MNIST digits or Fashion-MNIST (28 × 28 pixels), generating speckle patterns (256 × 256 pixels) that were subsampled via programmable SPD arrays. Three configurations were tested for MNIST and Fashion-MNIST: 7 × 7, 6 × 6, and 5 × 5 SPD arrays, where each detector integrated intensity over a 10 × 10 px region. This resulted in sparse sequences of 49, 36, or 25 intensity values per image. For both datasets, the 60,000 images were partitioned into training, validation, and test sets in an 8:1:1 ratio.

Remarkably, SMPD achieved 92%, 89.45%, and 83.28% recognition accuracy on the MNIST test dataset (6000 samples) for 7 × 7, 6 × 6, and 5 × 5 SPD arrays, respectively (Fig. 6b). The confusion matrix (Fig. 6c) reveals minimal misclassification trends, with errors predominantly occurring between morphologically similar digits (e.g., 3 ↔ 5, 4 ↔ 9). For the more visually complex Fashion-MNIST dataset, SMPD achieved test accuracies of approximately 84%, 82%, and 81% for each SPD configuration. As shown in the confusion matrix (Fig. 6c), misclassifications were concentrated among upper-body garments (such as shirts, T-shirts, pullovers, and coats), whose silhouettes exhibit high visual similarity; representative examples are highlighted in Fig. 6a.

These results confirm that spatially sparse, non-locally encoded speckle measurements encapsulate sufficient information for high-accuracy classification across diverse, non-orthogonal image sets, without resorting to pixel-dense sensors. By reducing sensor-pixel requirements to a low subsampling rate (e.g., 5 × 5 arrays vs. 28× 28 pixels, 3.19% sampling density), SMPD offers a computationally efficient, hardware-lean pathway to deploy optical classifiers in scenarios where pixel-dense cameras are impractical or where speed, power, and footprint are at a premium.

## Discussion

The success of SMPD in achieving high-accuracy OAM recognition with ultra-low sampling density stems from the intrinsic properties of speckle patterns formed in complex media. Unlike free-space propagation, scattering processes in MMFs redistribute input wavefront information across the entire speckle field, creating spatially redundant correlations. That is, the information projected onto a single speckle grain on the output plane is the superposition of the entire incident light field[32,33,40,41]. Such non-local encoding enables partial-field sampling to retain sufficient information for OAM retrieval, a principle validated by transmission matrix theory and prior work on scattering media[29,30].

By exploiting this redundancy, SMPD circumvents the need for high-resolution imaging, which has long constrained the scalability of OAM-based technologies[9,35,36]. While conventional methods[8,9,35,36] require dense pixel arrays (e.g., over 256 × 256 pixels) or full-field images to resolve OAM-dependent or other non-orthogonal features, SMPD achieves comparable accuracy with just 16 SPDs (0.024% sampling density). This efficiency is further enhanced by the ROAM-ANN neural network, which learns discriminative features from the multiplexed intensity sequences, even under photon-limited conditions. Compared to traditional free-space approaches, SMPD represents an efficient sensing paradigm with minimal data requirements, well-suited for complex or constrained environments.

Consequently, our findings indicate that the reliance on spatially resolved cameras for OAM recognition can be greatly alleviated. By harnessing speckle-intensity data from only a small number of SPDs, the proposed method enables high-speed, low-cost, single-shot detection suitable for a wide range of practical scenarios. Furthermore, its compatibility with single-pixel detectors extends the technique to spectral regimes (e.g., infrared, ultraviolet) where high-resolution cameras are prohibitively expensive or simply unavailable.

The distance between sampling points should exceed the speckle grain size, as neighboring points within a speckle pattern tend to exhibit high correlation. This is due to the nature of the normalized intensity correlation function, whose decay characterizes the speckle grain size. It means that points located within a single speckle grain are highly correlated and thus provide redundant intensity information. Conversely, sampling points spaced farther apart are more likely to capture independent modal interference patterns, enhancing the diversity of information available for classification. While the total information content may still be preserved within closely spaced regions when the scattering effect is sufficiently strong, a sparse sampling strategy is preferred to prevent overlapping within the correlation range. Moreover, scattered sampling proves robust across different regimes, performing well whether the speckle grains are large or small (see details in Supplementary Note 1.4).

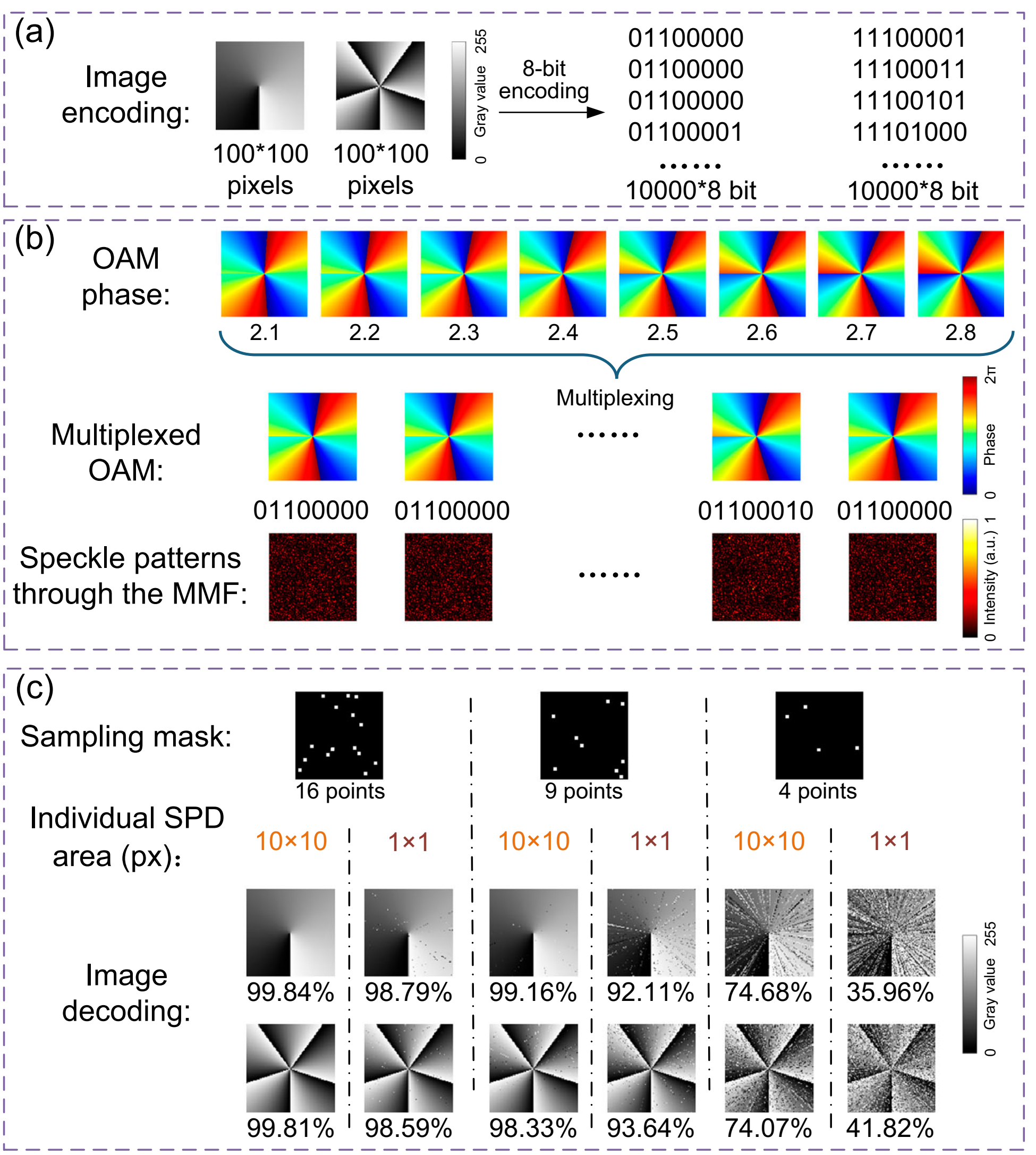

**Fig. 5 | High-capacity OAM-multiplexed optical communication via SMPD.** **a** Encoding process: Two spiral phase images (100 × 100 pixels, 8-bit depth) are encoded by using OAM bases ($\ell = 2.1$–$2.8$), and the input multiplexed is achieved by using DMD combined with Lee hologram encoding method. **b** Transmission and speckle generation: The multiplexed beams propagate through a 20-meter MMF, generating mode-scrambled speckle patterns at the output. **c** Decoding performance: Reconstructed images and error rates for three SPD configurations, 16 SPDs (10 × 10 px each) achieve <0.2% decoding error, while 4 SPDs retain 74% accuracy.

Although a single pixel is theoretically effective in each sampling region, in practice, individual pixels are susceptible to various sources of noise, such as mechanical vibrations (causing pixel displacement) or sensor-related noise (e.g., dark current). Sampling over a denser pixel region (ideally at least half a speckle grain) helps mitigate these effects by averaging out noise, thereby enhancing the reliability and stability of the measured intensity. In practice, this concern is further alleviated because commercial SPDs have far larger active areas and far more stable responsivities than the CCD-based "virtual SPDs" used in our systems.

A second consideration is the intrinsic sparsity of the speckle field: destructive interference leaves extensive zones of near-zero intensity. Expanding the sampled area within each SPD therefore increases the probability of intercepting bright speckle grains and reduces the risk of drawing data from uninformative regions, ultimately preserving the information content of the measurement embedded within the speckle pattern.

A pivotal insight from our experiments is that, while each SPD should subtend an area large enough to average pixel-level noise, the dominant performance driver remains the number and distribution of SPDs. This makes SMPD highly effective for classification, yet its sparse-sampling framework becomes information-limited in high-entropy tasks such as detailed image reconstruction. These challenges stem from the inherent trade-off between sampling sparsity and information capacity. The most direct remedy is to raise the sampling density, and two complementary strategies are emerging to do so without sacrificing speed or footprint: (i) ultracompact metasurface-based SPD arrays that multiply detector counts, and (ii) compressed-sensing algorithms that extract more information from the same measurements. Together, these approaches may extend SMPD to real-time, high-fidelity imaging through scattering media.

Beyond OAM recognition, our experiment results (Fashion-MNIST, MNIST classification, and axicon angle detection in Supplementary Note 7) demonstrate SMPD's versatility in decoding diverse optical information, from orthogonal to non-orthogonal domains. This generalizability positions SMPD as a universal platform for speckle-based sensing, with transformative potential for quantum communication[14], low-power LIDAR[17], and NLOS sensing[7,42]. Future work could explore hybrid frameworks combining SMPD with quantum-enhanced detectors[43], adaptive optics[44], or wavefront shaping[45], further bridging the gap between theoretical information redundancy and practical, resource-efficient photonic systems.

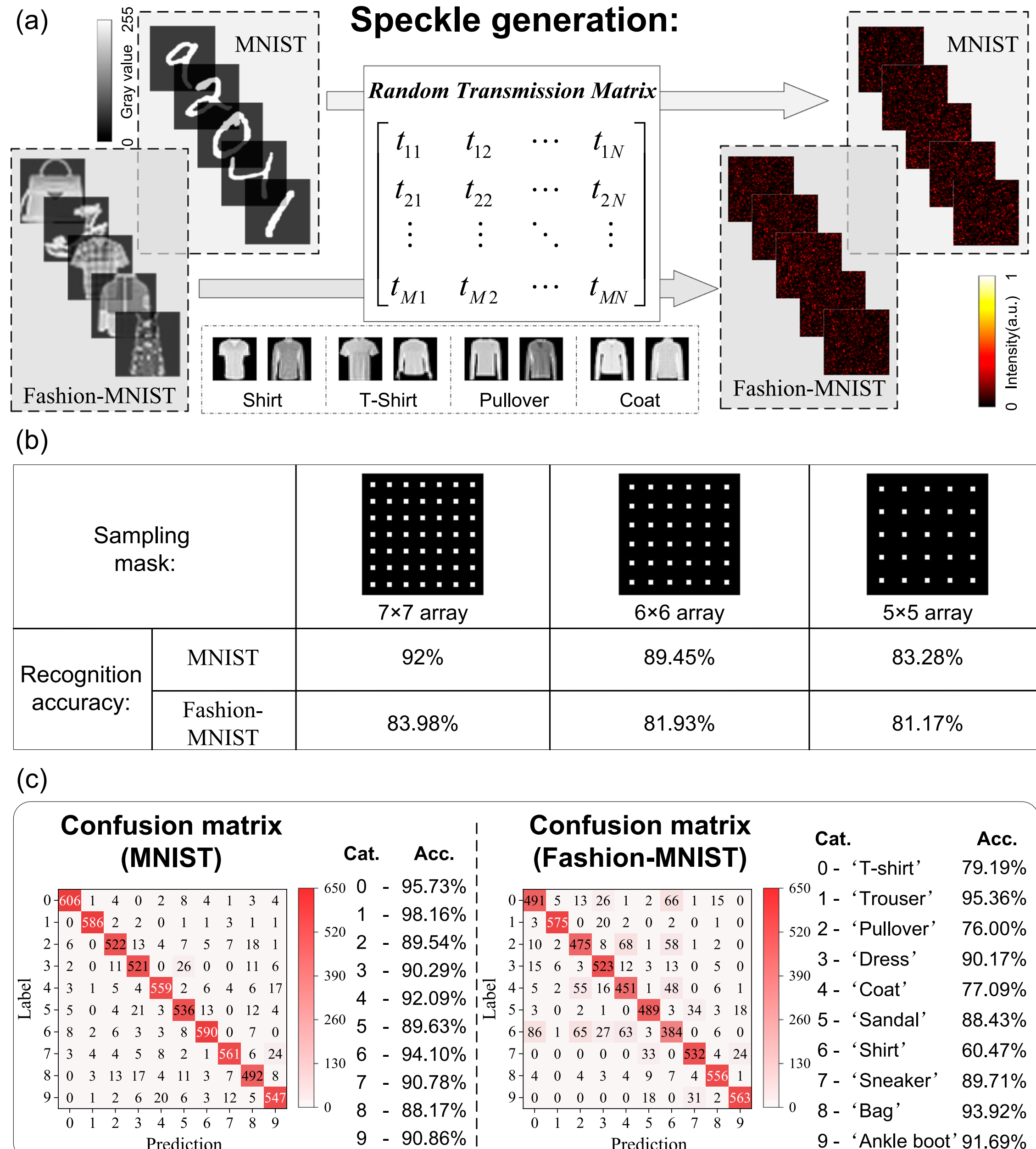

**Fig. 6 | Grayscale image and handwritten digits recognition using SMPD with different sampling masks. a** Simulated scattering transmission and sparse detection: Images from the MNIST (handwritten digits) and Fashion-MNIST (28 × 28 pixels) were numerically propagating through a scattering medium (random transmission matrix), generating speckle patterns (256 × 256 pixels). Programmable sampling masks (7 × 7, 6 × 6, or 5 × 5 SPD arrays) extracted mean intensity values from 10 × 10 px regions, emulating sparse single-pixel detection. b Performance metrics: Recognition accuracy for each SPD configuration on the MNIST and Fashion-MNIST test datasets (6000 samples each). For MNIST, SMPD achieved recognition accuracies of 92.00%, 89.45%, and 83.28% with 7 × 7, 6 × 6, and 5 × 5 arrays, respectively. For Fashion-MNIST, the corresponding accuracies are 83.98%, 81.93%, and 81.17%, respectively. **c** Confusion matrices: Classification results for MNIST and Fashion-MNIST datasets using the 7 × 7 SPD array (49 detectors). Abbreviations: Cat., category; Acc., accuracy.

In summary, we have demonstrated a speckle-driven framework, spatially multiplexed points detection (SMPD), that achieves efficient OAM recognition with unprecedented sampling efficiency. By exploiting the non-local correlations inherent to speckle patterns, SMPD extracts OAM information from sparse intensity measurements using 16 SPDs with a sampling density of 0.024%. This represents a 4096-fold improvement over conventional imaging-based methods, while maintaining >99% accuracy. The method's insensitivity to SPD spatial arrangements, combined with its compatibility with low-cost fast detectors, enables deployment in resource-constrained environments, from endoscopic imaging to high-speed optical communication. Experiments on OAM-multiplexed data transmission, spatiotemporally interleaved OAMs decoding, and successful classification of MNIST and Fashion-MNIST further validate SMPD's versatility and practicality. The synergy between sparse sampling and machine learning not only addresses the ill-posed challenges of MMF transmission but also establishes a blueprint for scalable photonic technologies.

Furthermore, by utilizing single-pixel detectors without spatial resolution, SMPD departs from conventional free-space detection schemes, offering a sensing modality that combines low data requirements with high discriminative power. Looking ahead, SMPD's modular design invites integration with advanced computational algorithms and compact detector arrays, promising breakthroughs in high-dimensional optical sensing, real-time quantum state tomography, and miniaturized medical diagnostics. By redefining the spatial sampling-information capacity trade-off, this work advances the frontier of optical technologies, enabling efficient light-matter interaction control in complex, scattering environments.

## Methods

### Speckle generation of orthogonally polarized vortex beams

A polarized femtosecond laser passes through a half-wave plate to adjust its polarization to 45°. A polarizing beam splitter (PBS) divides the 45° polarized beam into two orthogonally polarized paths. Each path is independently modulated by a spatial light modulator (SLM) to generate vortex beams with different OAMs. Then, the vortex beams are recombined by the PBS and coupled into an MMF using a 10× objective. At the distal end of the fiber, speckle patterns are collected by a 20× objective and recorded by a CCD camera (7.4 μm pixel size). (More details can be found in Supplementary Materials Note 2).

### Speckle generation of Spatiotemporally interleaved OAMs

Eight vortex beams carrying OAM values ($\ell = 2.1–2.8$) are sequentially modulated by a DMD illuminated with a continuous-wave laser. Each vortex beam is generated by Lee hologram[37] and temporally modulated by a distinct chronological mask, and subsequently coupled into an MMF through a 10× objective. The output speckle field is collected by a 20× objective lens and recorded by a CCD camera. The entire process is synchronized using a function generator, ensuring temporal alignment between DMD modulation and CCD exposure. During each speckle acquisition period, the CCD remains continuously exposed to integrate the intensity of the speckle field generated by the eight sequentially transmitted vortex beams.

### Multiplexed OAM transmission and speckle generation

A continuous-wave laser is expanded and directed onto the DMD, which sequentially displays the CGHs corresponding to the encoded gray values. The modulated beam is relayed through the 4 f system and coupled into an MMF using a collimator. The output field is collected by a 20× objective lens and recorded by a CCD camera. In the data acquisition process, the DMD sequentially loads CGHs corresponding to grayscale values ranging from 0 to 255, each repeated ten times to improve signal stability. Two spiral phase images of 100 × 100 pixels are encoded into 20,000 CGHs, forming the testing dataset. To further mitigate environmental noise, all CGHs associated with grayscale values 0–255 are re-captured in ten repetitions after image transmission. In total, 5120 intensity-label pairs constitute the training dataset, while 20,000 CGHs serve as the testing set for evaluating multiplexed-OAM transmission (More details can be found in Supplementary Materials Note 5).

### Speckle transformation of MNIST and Fashion-MNIST datasets

To numerically simulate the optical speckle transformation of standard image datasets, we modeled a transmission process through a random scattering medium using a complex-valued transmission matrix. Each 28 × 28 pixels image from the MNIST[38] or Fashion-MNIST[39] datasets was first vectorized into a column vector and normalized to the range [0, 1]. A random complex-valued transmission matrix **T** (with row of $28^2$ rows and column of $256^2$) was then generated to represent the scattering process. The resulting out field (256 × 256) was obtained by multiplying **T** with the input image vector, treating the normalized values as the input amplitudes. The speckle intensity pattern was calculated as the squared magnitude of this complex field. Finally, the speckle intensity was partitioned into a grid of localized regions (5 × 5, 6 × 6, or 7 × 7), and the summed intensity within each region was used as a representative feature vector.

### Neural network architecture and training

A dual-path neural network (ROAM-ANN) is designed to extract both local and global correlations within the intensity sequences. One path employs multiple 1D convolutional layers to capture spatially associated features, while the other uses fully connected layers to extract distributed sequence characteristics. The features from both paths are concatenated and fed into output modules with softmax activation. The network was implemented in TensorFlow 2.6 and trained with the Adam optimizer using binary cross-entropy loss. All experiments were performed on an NVIDIA P620 (More details can be found in Supplementary Note 4).

## Data availability

All relevant data are available from the corresponding author upon request.

## Code availability

All relevant code is available from the corresponding author upon request.

## Acknowledgements

Z. C. acknowledges support by the National Natural Science Foundation of China (Grant no. 62375092) and Key Project of Natural Science Foundation of Fujian Province (No. 2023J02020). V. R. acknowledges support by the Fundamental Research Funds for the Central Universities (ZQN-1205). P. L. acknowledges support by the National Natural Science Foundation of China (Grant nos. 81930048 and 82330061), Hong Kong Research Grant Council (15125724 and C7074-21G), Guangdong Science and Technology Commission (2019BT02X105), Shenzhen Science and Technology Innovation Commission (JCYJ20220818100202005) and Hong Kong Polytechnic University (P0038180, P0039517, P0043485, P0045762, and P0049101).

## Author contributions

Z.W., Z.C., H.L., and P. L. conceived the idea; Z.W. and H.L. conducted the simulations and experiments. Z.W. and H.L. analyzed the data and results and improved the algorithm performance with assistance from T.Z., Q.Z., H.H.L. and Z.Y., Z.W., H.L., V.R., Z.C. and P.L. wrote and revised the manuscript. Z.W., H.L., Z.Y., Z.C. and P.L. prepared the Supplementary Materials. Z.C., X.Y., J.P. and P.L. supervised the project. Z.W., H.L., V.R., Z.Y., H.H.L., J.P., Z.C., X.Y and P.L. proofread and optimized the manuscript.

## Competing interests

The authors declare no competing interests.

## Additional information

**Supplementary information** The online version contains supplementary material available at https://doi.org/10.1038/s41467-025-66074-3.

**Correspondence** and requests for materials should be addressed to Ziyang Chen, Xiaocong Yuan or Puxiang Lai.

**Peer review information** *Nature Communications* thanks Isaac Nape and the other anonymous reviewer(s) for their contribution to the peer review of this work. A peer review file is available.